\documentclass[conference]{IEEEtran}
\IEEEoverridecommandlockouts

\usepackage{cite}
\usepackage{amsmath,amssymb,amsfonts}
\usepackage{acronym}
\usepackage{subfigure} 
\usepackage{multirow}
\usepackage{booktabs}
\usepackage{algorithmic}
\usepackage{graphicx}
\usepackage{textcomp}
\usepackage{xcolor}

\def\BibTeX{{\rm B\kern-.05em{\sc i\kern-.025em b}\kern-.08em
    T\kern-.1667em\lower.7ex\hbox{E}\kern-.125emX}}

\input{AcronymList}
\begin{document}

\title{Cyclic Prefix (CP) Jamming Against Eavesdropping Relays in OFDM Systems 
\thanks{This work has been submitted to the IEEE for possible publication. Copyright may be transferred without notice, after which this version may
no longer be accessible.}
}

\author{
\IEEEauthorblockN{Muhammad~Sohaib~J.~Solaija\IEEEauthorrefmark{1}, Haji~M.Furqan\IEEEauthorrefmark{1}, Zekeriyya~Esat~Ankaral{\i}\IEEEauthorrefmark{2}\IEEEauthorrefmark{3} and H\"{u}seyin Arslan\IEEEauthorrefmark{1}\IEEEauthorrefmark{2}}\\
\IEEEauthorblockA{\IEEEauthorrefmark{1}Department of Electrical and Electronics Engineering, Istanbul Medipol University, Istanbul, 34810 Turkey\\}
\IEEEauthorblockA{\IEEEauthorrefmark{2}Department of Electrical Engineering, University of South Florida, Tampa, FL 33620 USA}
\IEEEauthorblockA{\IEEEauthorrefmark{3}Maxlinear, Inc., Carlsbad, CA 92008 USA\\
Emails: solaija@ieee.org, haji.madni@medipol.edu.tr, zekeriyya@mail.usf.edu, huseyinarslan@medipol.edu.tr
}}

\maketitle

\begin{abstract}
Cooperative communication has been widely used to provide spatial diversity benefits for low-end user equipments, especially in ad hoc and wireless sensor networks. However, the lack of strong authentication mechanisms in these networks leaves them prone to eavesdropping relays. In this paper, we propose a secure orthogonal frequency division multiplexing (OFDM) transmission scheme, where the destination node transmits a jamming signal over the cyclic prefix (CP) duration of the received signal. Simulation results verify that as long as at least a part of the jamming signal falls to the actual data portion of the eavesdropping relay, it spreads through all the data symbols due to the fast Fourier transformation (FFT) operation, resulting in degraded interception at the eavesdropper. 
\end{abstract}


\section{Introduction}
\label{Sec:Intro}
\par Wireless communication has become an increasingly important part of human existence, a reality re-emphasized by the recent pandemic \cite{saeed2020wireless}. The ubiquitous presence of wireless signals brings about significant advantages in terms of connectivity for communication and environmental awareness. However, this comes at the potential cost of users' data privacy and security \cite{furqan2021wireless}. In the current work, we focus on this problem in the context of cooperative communication in the presence of untrusted relays.

\par \Ac{MIMO} is considered to be an essential enabler for \ac{5G} (and even Wi-Fi) and beyond wireless networks \cite{IEEE_specturm}. However, services such as \ac{mMTC} and \ac{IoT} cannot always ensure the availability of hardware capabilities that can support \ac{MIMO} operation. This led to the emergence of \textit{cooperative communication} paradigm around the turn of the century, where the aim is to improve the \ac{QoS} of the users through cooperation. Cooperative communication, while generally applicable to different network realizations, is more suited to \textit{ad hoc} and \acp{WSN} compared to the cellular systems \cite{nosratinia2004cooperative}. Unlike the latter, the former two systems lack a sophisticated authentication mechanism. This may lead to violation of a user's data confidentiality by the very nodes that are expected to improve the communication via spatial diversity. Accordingly, there exists a plethora of approaches for \ac{PLS} in cooperative communication, which can be broadly categorized into \textit{cooperative relaying}, \textit{cooperative jamming} and their hybrid combinations. Both these approaches usually comprise of two phases, where the first one is broadcast of information from the source. In the second step, the appropriate relay forwards the message to the destination. In the case of relaying, the selection of the helper (relay) is done to degrade the eavesdropper's interception capability. In the case of jamming, the source itself, the destination, or any other helper node transmits noise-like signals such that the eavesdropper cannot hear the message \cite{pahuja2019cooperative}.

\par Despite the clear advantage in terms of degraded eavesdropping capability \cite{jameel2018comprehensive}, the cooperative jamming techniques have various limitations. For instance, in the case of friendly jamming the presence of a trustable node is necessary \cite{zhang2012physical}. Furthermore, the jamming signal from this helper can also degrade the destination's performance. Destination-based jamming techniques resolve this issue by exploiting their prior knowledge of the jamming signal. However, they require full-duplexing capabilities to achieve spatial diversity \cite{zhao2018secrecy}. In both cases, jamming signals are to be transmitted throughout the first phase (where the source broadcasts the signal) of cooperative communication, which is unsuitable for power-limited devices. To this end, we propose the design of a jamming signal which is only transmitted for part of the broadcast phase, yet its effect is spread throughout the signal duration by exploiting the properties of \ac{OFDM} signals and \ac{FFT} operation. Since the jamming signal is only transmitted during the \ac{CP} part at the destination, it also eliminates the need for full-duplexing capability. 

\par The rest of this article is organized as follows. Section \ref{Sec:SysModel} describes the general system model for \ac{AAF} cooperative relaying. The proposed approach for \ac{CP} jamming is discussed in Section \ref{sec:Proposed}. Simulation results and related discussions are covered in \ref{sec:Results} while Section \ref{sec:Conclusion} concludes the work and provides some future directions. 

\section{System Model}
\label{Sec:SysModel}
\par
\begin{figure}[t!]
    \centering
    \subfigure[]{\includegraphics[width=0.9\columnwidth]{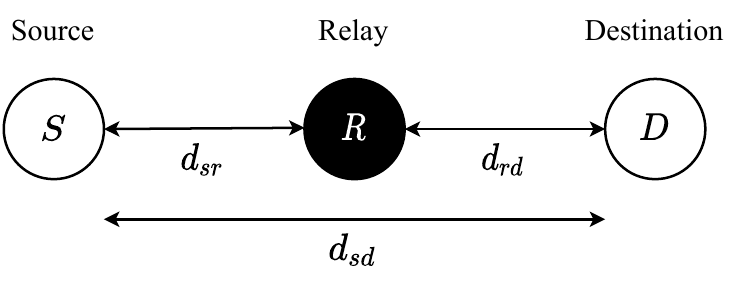}\label{fig:SysModel}} 
    \subfigure[]{\includegraphics[width=0.9\columnwidth]{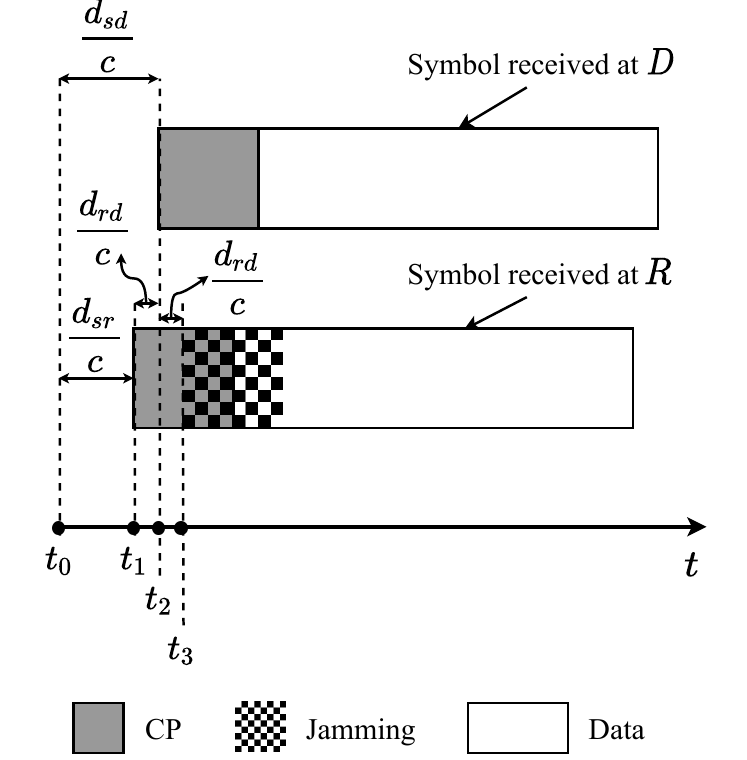}\label{fig:Proposed}} 
    \caption{Illustration of the proposed approach in a relaying scenario. (a) A typical dual-hop relay aided system comprising of a single source, relay and destination each. (b) The user signal is transmitted from source ($S$) at time $t_0$, reaching the relay ($R$) and destination ($D$) at times $t_1$ and $t_2$, respectively, while the jamming signal is transmitted by $D$ at $t_2$ and reaches $R$ at $t_3$.}
    \label{fig:eqqAPP}
\end{figure}

 Without loss of generality, we assume a dual-hop half-duplex relay-aided system with \ac{AAF} protocol \cite{liu2009cooperative}. As shown in Fig. \ref{fig:SysModel}, it consists of a source that wants to communicate with a destination in the presence of an untrusted relay, where each node consists of a single antenna. Here, the goal is to use the advantages provided by the relay without letting it decode any information. The system is based on \ac{OFDM} modulation with $N$ subcarriers. The channels corresponding to source-destination ($H_{sd}$), source-relay ($H_{sr}$), and relay-destination ($H_{rd}$) links are assumed to be slowly varying Rayleigh fading with $L$ exponentially decaying taps. A \ac{CP} of length $L$ is introduced to convert the multipath frequency channel into flat fading subchannels. 

\par At the transmitter, the frequency domain \ac{OFDM} symbols $\mathbf{{S}}=\begin{bmatrix} S_{0} & S_{1} &...& S_{N-1}\end{bmatrix}^{\textrm T} \in \mathbb{C}^{[N \times 1]}$ are passed through the \ac{IFFT} process as $s[n]=\sum_{k=0}^{N-1} S_{k} \exp^{\frac{j 2 \pi nk}{N}}$ and a \ac{CP} of length $L$ is added. The transmitted signal with \ac{CP} can be written as 
$\mathbf{{s}}=[s[N-L+1], \dots, s[N-1], s[0], s[1], \dots, s[N-L], s[N-L+1], \dots,  s[N-1] ]$. Finally, the source node broadcasts the information to relay and destination. The received signal at the destination in the frequency domain on $k$-th subcarrier can be represented by
\begin{equation}
\begin{aligned}
\label{eq:Y_sd}
Y_{sd}(k) &= \sqrt{P_1} H_{sd}(k) S(k) + V_{sd}(k) ,\\k &= 1,\dots, K;~n = 1, \dots, N,
\end{aligned}
\end{equation}
where $H_{sd}(k)$ is the \ac{CFR} of the $k$-th subcarrier between source and destination, the transmitted power by the source is represented by $P_{1}$, $S(k)$ is the symbol transmitted by source node on $k$-th subcarrier, and $V_{sd}(k)$ represents \ac{AWGN} on the $k$-th subcarrier with variance $N_{0}/2$. Similarly, the received signal at relay is given by
\begin{equation}
\begin{aligned}
\label{eq:Y_sr}
Y_{sr}(k) &= \sqrt{P_1} H_{sr}(k) S(k) + V_{sr}(k),
\end{aligned}
\end{equation}
where $H_{sr}(k)$ is the \ac{CFR} of the $k_{th}$ subcarrier for source-relay and $V_{sr}(k)$ represents the \ac{AWGN}. The relay node will normalize the received signal before retransmission by the factor  
\begin{equation}
\beta (k)=\sqrt \frac{1}{P_1 H_{sr}(k)+ N_{0}}.
\end{equation}

Finally, the received signal at destination from the relay can be given as 
\begin{equation}
\begin{aligned}
\label{eq:Y_rd}
Y_{rd}(k) &= \sqrt{P_2} H_{rd}(k) \Big(\beta (k) Y_{sr}(k))\Big) + V_{rd}(k), 
\end{aligned}
\end{equation}
where $H_{rd}(k)$ is the frequency domain channel attenuation on the $k$-th subcarrier between relay and destination, the transmitted power is represented by $P_{2}$, and  $V_{rd}(k)$ shows  \ac{AWGN} on the $k$-th subcarrier. The receiver will finally combine $Y_{rd}(k)$ signal with $Y_{sd}(k)$ given in (\ref{eq:Y_sd}) by applying \ac{MRC} and decode the information \cite{seo2006maximum}.

\section{Proposed Approach}
\label{sec:Proposed}

\par This section presents the details of the proposed algorithm. The basic idea here is to devise a method that enables us to utilize the advantages of the untrusted relay while keeping the information secure from being eavesdropped on by the same node. Moreover, we want to achieve this goal without using an external helper or full-duplex receiver. It should be noted that in the case of half-duplex destination, diversity gain is nullified since the destination transmits jamming signal throughout the first phase rendering it incapable of receiving the broadcast copy of the signal. The proposed approach, on the other hand, provides the required spatial diversity obtained from the signals received in both phases while ensuring that the relay is unable to intercept the data signal. This is achieved by the transmission of jamming signal over the \ac{CP} duration of the received \ac{OFDM} signal in the first phase.

The proposed algorithm has two phases.
\begin{itemize}
    \item \textbf{Phase-1}: Source transmits a communication signal while destination sends a jamming signal during a portion of signal duration.
    \item \textbf{Phase-2}: The relay forwards a signal that it receives during phase-1 to the destination for final combining and decoding at the destination. 
\end{itemize}
\noindent The following text provides details of the aforementioned phases of the proposed approach.
\subsection{Phase-1}

\par In phase-1, the source node broadcasts an \ac{OFDM} signal to destination and untrusted relay over multi-path Rayleigh fading channel. The received signals at destination and relay are given in (\ref{eq:Y_sd}) and (\ref{eq:Y_sr}), respectively. The relay forwards the received signals using \ac{AAF} protocol to the destination in phase-2 to enhance the reliability of communication. However, there is a possibility that the relay may try to intercept and decode the data signal for malicious purposes. To ensure reliable communication while ensuring security from the untrusted relay, we propose that the destination transmits the jamming signal over the \ac{CP} duration of the received signal right before receiving the \ac{OFDM} signal. Even though the jamming signal overlaps with only a portion of the time-domain transmitted signal, its effect spreads over all the data-containing subcarriers once the \ac{FFT} process is applied at the relay. This received signal at the relay, containing the jamming signal arriving from the destination is given by

\begin{equation}
\begin{aligned}
\label{eq:y_sr_hat}
\hat{Y}_{sr}(k) &= \sqrt{P_1} H_{sr}(k) S(k) + V_{sr}(k) \\&+\sqrt{P_j} H_{rd}(k) J(k)+ V_{rd}(k),
\end{aligned}
\end{equation}
where $J(k)$ is the jamming signal and the power of the jamming signal transmitted by the destination is represented by $P_{j}$. It should be noted that (\ref{eq:y_sr_hat}) contains the additional jamming term compared to (\ref{eq:Y_sr}), hence the term $\hat{Y}_{sr}(k)$ is used instead of ${Y}_{sr}(k)$.

\par Figure \ref {fig:Proposed} provides a temporal illustration of phase-1 and how the signals arrive at both destination and relay. The transmission from the source initiates at time $t_0$. The signal arrives at relay and destination nodes at times times $t_1$ and $t_2$, respectively. The destination then transmits the jamming signal, equal in length to the \ac{CP} itself and represented by a checkered block in the figure, immediately upon receiving its signal. The jamming signal reaches the relay at $t_3$. The starting point of the jammed portion of the relay’s signal in time can be calculated as
\begin{equation}
t_d = t_3 - t_1 = \frac{d_{sd}+d_{rd}-d_{sr}}{c},
\end{equation}
where $d_{sd}$ indicates the distance between source and destination, $d_{sr}$ is the distance between source and relay, and $d_{rd}$ is the distance between relay and destination. As mentioned earlier, as long as part of the jamming signal overlaps with the data portion at the relay, its decoding capability is degraded. The overlap itself is ensured by the propagation delay between the transceivers. 

\subsection{Phase-2}
\par In phase-2, the relay transmits the signal to the destination after employing \ac{AAF} protocol. The received signal at the destination from the relay without jamming is given by (\ref{eq:Y_rd}). Replacing ${Y}_{sr}(k)$ with $\hat{Y}_{sr}(k)$ in (\ref{eq:Y_rd}) and expanding it gives us
\begin{equation}
\begin{aligned}
\hat{Y}_{rd}(k) &= \sqrt{P_2} H_{rd}(k)\big(\beta(k) \sqrt{P_1} H_{sr}(k) S(k)\\& +  V_{sr}(k)+\sqrt{P_j} H_{rd}(k) J(k)+ V_{rd}(k)\big).\\ 
\end{aligned}
\end{equation}

\par The receiver will first remove the jamming signal from $\hat{Y}_{rd}(k)$. Since the jamming signal is known at the destination and \ac{CP} is discarded in the receiver anyway, the jamming signal does not affect the signal of the legitimate receiver. However, some reflected components of the jamming signal may leak into the actual data and cause a self-interference. In this study, we assume that the receiver is able to cancel that interference by estimating the channel around itself. Afterwards, it employs \ac{MRC} \cite{seo2006maximum} to combine the resultant signal with the $Y_{sd}(k)$ that it received during phase-1.  

\par Here it is important to highlight that unlike conventional \ac{PLS} techniques \cite{hamamreh2018classifications}, the proposed method does not require channel state information at the transmitter \cite{ankarali2016channel}. This renders the proposed approach suitable for scenarios where the transmitter/source node has limited capabilities in terms of computation, channel estimation, etc.

\begin{table}[t!]
\centering \renewcommand{\arraystretch}{1.25}
\caption{Simulation parameters and assumptions} \label{tab:sim_parameters}
\resizebox{0.9\columnwidth}{!}{%
\begin{tabular}{|l|l|}
\hline
\textbf{Parameter}      & \textbf{Value}                \\ \hline
Simulation environment             & Urban macro                   \\ \hline
Carrier frequency $(f_c)$     & 2 GHz                         \\ \hline
Shadow fading standard deviation $(\sigma)$        & 4 dB               \\ \hline
Source/relay transmit power ($P_{Tx}$)  & 23 dBm                        \\ \hline
Noise power density $(N_0)$     & -174 dBm/Hz                    \\ \hline
Source-destination distance     & 1000 m 
\\ \hline
Source-relay distance     & (100, 200, ... , 900) m 
\\ \hline
CIR length ($L$)      & 32 
\\ \hline
Modulation      & QPSK
\\ \hline
FFT Size ($N$)      & 256
\\ \hline
\end{tabular}%
}
\end{table}

\section{Simulation Results and Discussion}
\label{sec:Results}
\par As mentioned earlier, this work considers the presence of a single antenna source, destination, and relay nodes in an urban macro environment. As the ratio between received data and jamming signals depends heavily on the position of the relay, the following pathloss model is used \cite{3GPP_38_901}:
\begin{equation}
\label{pathloss}
    PL = 22\log_{10}(d)+20\log_{10}(f_c)+28+\sigma,
\end{equation}
where $d$ is the distance between the transceivers in meters, $f_c$ is the carrier frequency in GHz, and $\sigma$ represents the shadow fading modeled as a zero-mean log-normal distribution. The source and destination nodes are assumed to be $1$km apart, with the relay occupying possible positions at multiples of $100$m from them. The transmission power of both source and relay is selected as $23$dBm, which is typical of a cellular \ac{UE}. Each link experiences a slow Rayleigh fading channel, with an exponentially decaying power delay profile containing a \ac{CIR} of $L = 32$ taps, which is also the length of the \ac{CP} used (unless mentioned otherwise). Perfect synchronization and channel estimation is assumed for all links. Noise \ac{PSD}, $N_0$, is taken to be $-174$dBm/Hz. The simulation results are averaged over $5000$ \ac{OFDM} blocks, where each block contains $256$ \ac{QPSK} symbols. A summary of these simulation parameters is provided in Table \ref{tab:sim_parameters}.

\begin{figure}[t!]
    \centering
    \includegraphics[width=1\columnwidth]{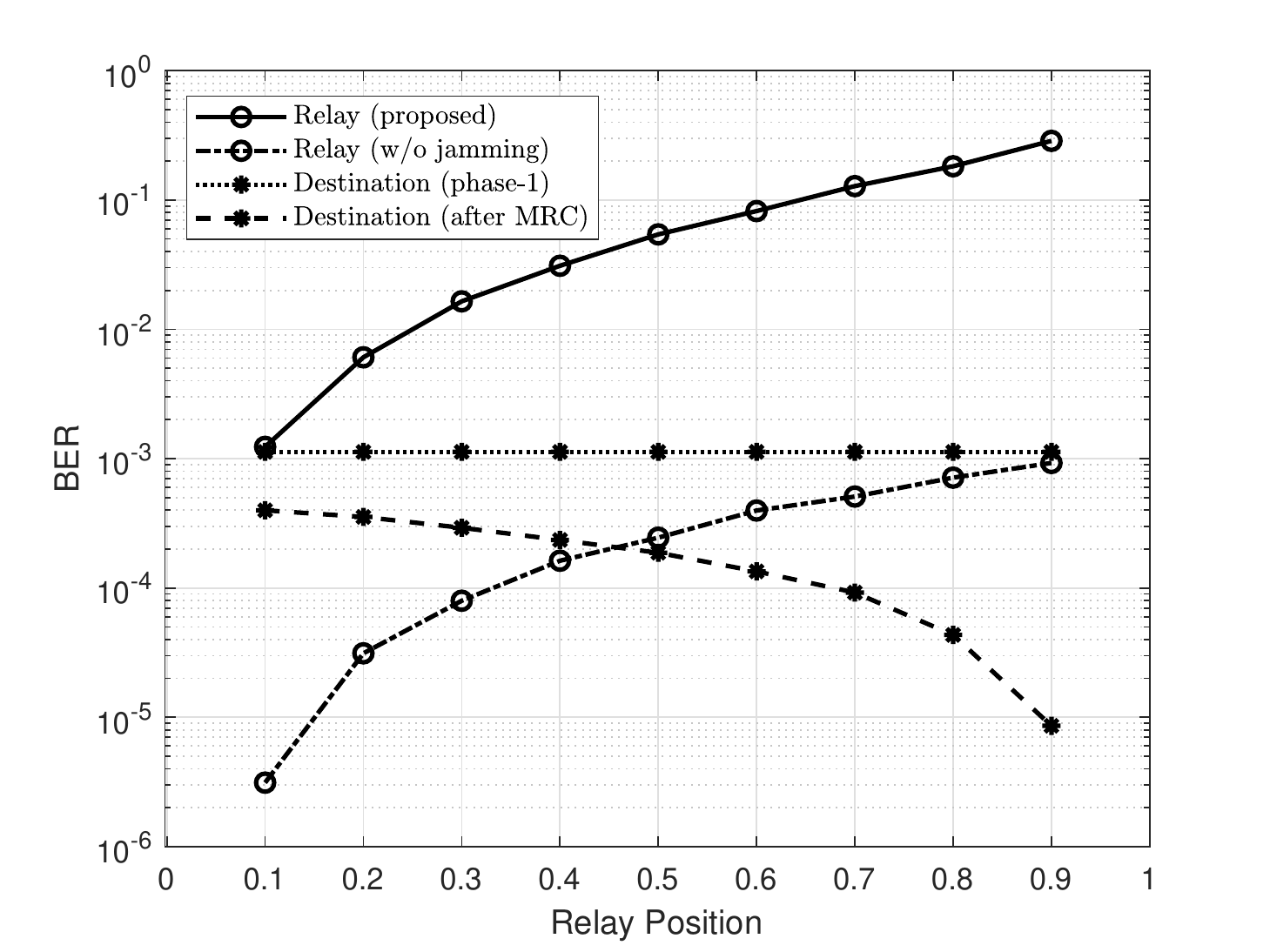}
    \caption{BER performance comparison of relay and destination. Relay's performance is seen to be degraded severely with the proposed CP jamming scheme.}
    \label{fig:rel_dest_comparison}
\end{figure}

\par Figure \ref{fig:rel_dest_comparison} shows the performance of relay and destination nodes in terms of \ac{BER}. This is used to represent the \textit{security gap}, which is quantified by the gap in error rates (bit, symbol, packet, etc.) of the legitimate and illegitimate receivers \cite{hamamreh2018classifications}. The horizontal axis represents the position of the relay, where zero is the location of the source itself, while $1$ represents the location of the destination. For this simulation, both data and jamming signals are assumed to have the same power levels ($23$ dBm). The two curves for relay represent the two cases of normal relay operation (without jamming) and with the proposed jamming approach, respectively, while the curves for destination represent its performance in phase-1 (where it only receives the broadcast signal from the source) and when it applies \ac{MRC} to the two copies received via source and relay. It can be seen that the relay's performance is significantly better than the destination when neither jamming nor combining is performed. This is simply due to the physical closeness of the relay to the source, and the consequent decrease in pathloss. It is shown in \cite{chen2009multi} that distance-based cooperative protocol selects the relay closest to the destination as the optimal one. Our results also re-iterate this, as the performance of destination after \ac{MRC} is significantly improved when the relay moves closer to it. The proposed algorithm also increases the \ac{BER} of the untrusted relay at this point, leading to increased security gap.

\begin{figure}[t!]
    \centering
    \includegraphics[width=1\columnwidth]{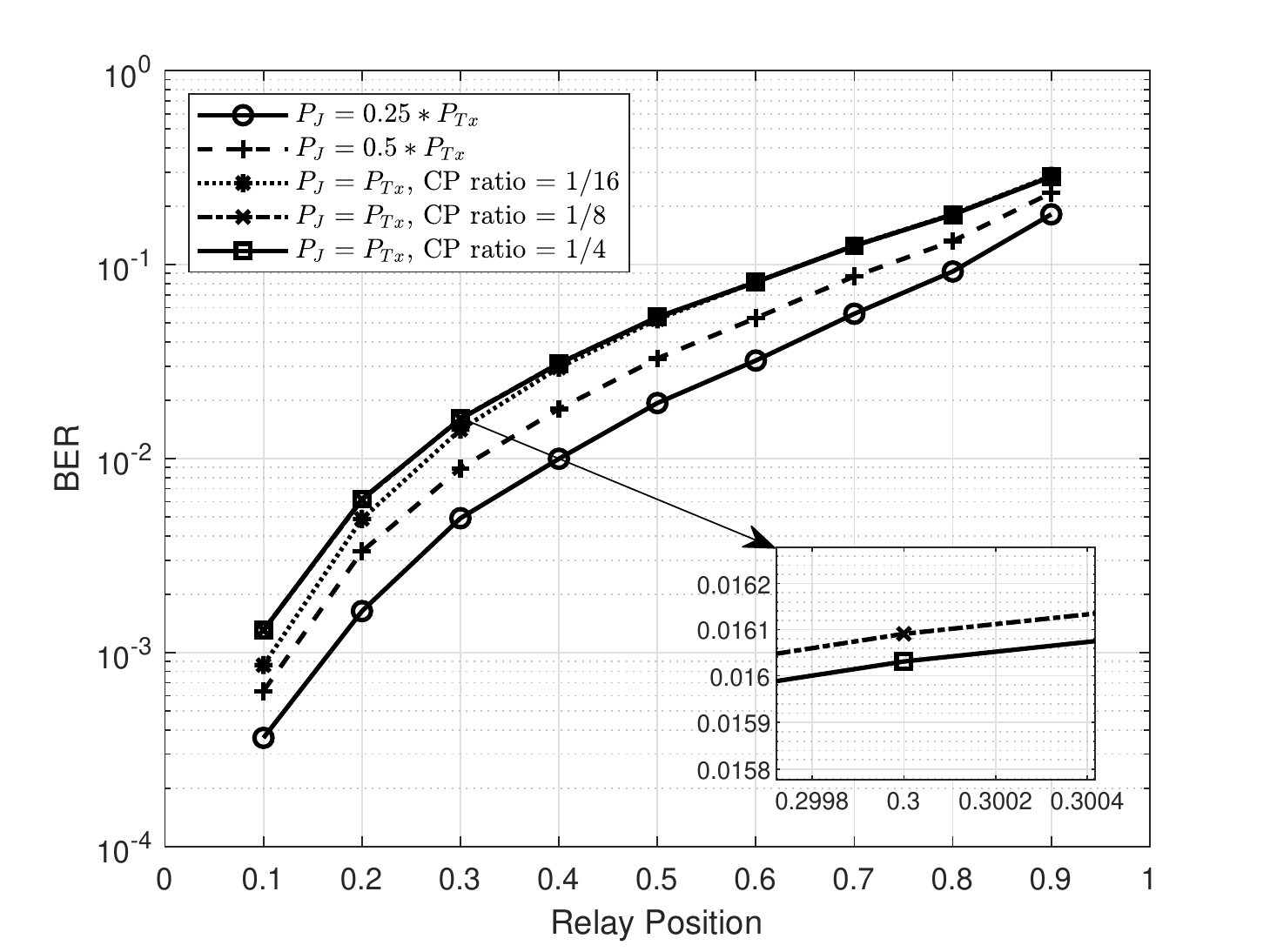}
    \caption{Effect of jamming power and \ac{CP} duration.}
    \label{fig:CP_Pjam}
\end{figure}

\par The interception capability of the eavesdropping relay depends on the jamming signal received. Accordingly, we look at the effect of different power levels of the jamming signals ($P_J$) as well as its different durations, as shown in Fig. \ref{fig:CP_Pjam}. For the former, we have considered three cases, i.e., $P_J = [0.25, 0.5, 1]*P_{Tx}$. It can be seen that variation in $P_J$ has a clearly distinguishable effect on the relay's \ac{BER}. For instance, when the relay is at the middle of the source and destination nodes, its \ac{BER} goes from $0.019$ to $0.05$ as $P_J$ increases from $0.25*P_{Tx}$ to $P_{Tx}$. To see the effect of the length of the jamming signal, we looked at three cases where both the length of \ac{CP} and jamming were varied, i.e., \ac{CP} ratios of $1/16, 1/8, 1/4$ were used (which were kept equal to \ac{CIR} itself). It is observed that the length of the jamming signal itself has no visible effect on the \ac{BER} performance. This observation also encourages us to use the minimum possible \ac{CP} length in the system without compromising on the security performance of the proposed algorithm.

\section{Conclusion and Future Directions}
\label{sec:Conclusion}
Wireless communication has become exceedingly important in our daily lives. However, one of the major concerns in future networks is the privacy of user data. To this end, this work tries to address the eavesdropping problem caused by untrusted relays. The proposed approach involves the transmission of a jamming signal from the destination in the interval while it receives the \ac{CP} part of the signal. Unlike the conventional solutions which require jamming signals to be transmitted throughout the broadcast phase, the proposed method only requires the signal to be transmitted for a fraction of it.

\par It should be noted that even though the current work focuses on the cooperative communication scenario, specifically untrusted relays, the concept presented in this work is much more general and can be applied to the eavesdropping problem in any \ac{OFDM} system. Moreover, in such systems the jamming signal may also corrupt the (blind) synchronization attempted by any illegitimate receiver. Further work can also be carried out to optimize the relay placement and jamming power allocation to maximize the security gap in cooperative systems. Moreover, the proposed technique can be used for secure key exchange between the legitimate nodes in the presence of an untrusted relay. Furthermore, the proposed algorithm can also be applied in cases when there is no direct path between source and destination.

\section*{Acknowledgment}
This work was supported in part by the Scientific and Technological Research Council of Turkey (TUBITAK) under Grant No. 120C142. The work of Haji M. Furqan was supported by HISAR Lab at TUBITAK BILGEM, Gebze, Turkey.

\bibliographystyle{IEEEtran}

\begin{thebibliography}{99}
\providecommand{\url}[1]{#1}
\csname url@samestyle\endcsname
\providecommand{\newblock}{\relax}
\providecommand{\bibinfo}[2]{#2}
\providecommand{\BIBentrySTDinterwordspacing}{\spaceskip=0pt\relax}
\providecommand{\BIBentryALTinterwordstretchfactor}{4}
\providecommand{\BIBentryALTinterwordspacing}{\spaceskip=\fontdimen2\font plus
\BIBentryALTinterwordstretchfactor\fontdimen3\font minus
  \fontdimen4\font\relax}
\providecommand{\BIBforeignlanguage}[2]{{%
\expandafter\ifx\csname l@#1\endcsname\relax
\typeout{** WARNING: IEEEtran.bst: No hyphenation pattern has been}%
\typeout{** loaded for the language `#1'. Using the pattern for}%
\typeout{** the default language instead.}%
\else
\language=\csname l@#1\endcsname
\fi
#2}}
\providecommand{\BIBdecl}{\relax}
\BIBdecl

\bibitem{saeed2020wireless}
N.~Saeed, A.~Bader, T.~Y. Al-Naffouri, and M.-S. Alouini, ``When wireless
  communication faces covid-19: Combating the pandemic and saving the
  economy,'' \emph{arXiv preprint arXiv:2005.06637}, 2020.

\bibitem{furqan2021wireless}
H.~M. Furqan, M.~S.~J. Solaija, H.~T{\"u}rkmen, and H.~Arslan, ``{Wireless
  communication, sensing, and REM: a security perspective},'' \emph{IEEE Open
  Journal of the Communications Society}, vol.~2, pp. 287--321, 2021.

\bibitem{IEEE_specturm}
\BIBentryALTinterwordspacing
{Amy Nordum, Kristen Clark and IEEE Spectrum Staff}, \emph{Everything You Need
  to Know About 5G}, accessed Aug. 3, 2020. [Online]. Available:
  \url{https://spectrum.ieee.org/video/telecom/wireless/everything-you-need-to-know-about-5g}
\BIBentrySTDinterwordspacing

\bibitem{nosratinia2004cooperative}
A.~Nosratinia, T.~E. Hunter, and A.~Hedayat, ``Cooperative communication in
  wireless networks,'' \emph{IEEE communications Magazine}, vol.~42, no.~10,
  pp. 74--80, 2004.

\bibitem{pahuja2019cooperative}
S.~Pahuja and P.~Jindal, ``Cooperative communication in physical layer
  security: Technologies and challenges,'' \emph{Wireless Personal
  Communications}, vol. 108, no.~2, pp. 811--837, 2019.

\bibitem{jameel2018comprehensive}
F.~Jameel, S.~Wyne, G.~Kaddoum, and T.~Q. Duong, ``A comprehensive survey on
  cooperative relaying and jamming strategies for physical layer security,''
  \emph{IEEE Communications Surveys \& Tutorials}, vol.~21, no.~3, pp.
  2734--2771, 2018.

\bibitem{zhang2012physical}
R.~Zhang, L.~Song, Z.~Han, and B.~Jiao, ``Physical layer security for two-way
  untrusted relaying with friendly jammers,'' \emph{IEEE Transactions on
  Vehicular Technology}, vol.~61, no.~8, pp. 3693--3704, 2012.

\bibitem{zhao2018secrecy}
R.~Zhao, X.~Tan, D.-H. Chen, Y.-C. He, and Z.~Ding, ``Secrecy performance of
  untrusted relay systems with a full-duplex jamming destination,'' \emph{IEEE
  Transactions on Vehicular Technology}, vol.~67, no.~12, pp. 11\,511--11\,524,
  2018.

\bibitem{liu2009cooperative}
K.~R. Liu, A.~K. Sadek, W.~Su, and A.~Kwasinski, \emph{Cooperative
  communications and networking}.\hskip 1em plus 0.5em minus 0.4em\relax
  Cambridge university press, 2009.

\bibitem{seo2006maximum}
B.-S. Seo, S.-G. Choi, and J.-S. Cha, ``{Maximum ratio combining for OFDM
  systems with cochannel interference},'' \emph{IEEE Transactions on Consumer
  Electronics}, vol.~52, no.~1, pp. 87--91, 2006.

\bibitem{hamamreh2018classifications}
J.~M. Hamamreh, H.~M. Furqan, and H.~Arslan, ``{Classifications and
  applications of physical layer security techniques for confidentiality: A
  comprehensive survey},'' \emph{IEEE Communications Surveys \& Tutorials},
  vol.~21, no.~2, pp. 1773--1828, 2018.

\bibitem{ankarali2016channel}
Z.~E. Ankaral{\i}, M.~H. Y{\i}lmaz, M.~Hafez, and H.~Arslan, ``Channel
  independent physical layer security,'' in \emph{IEEE 17th Annual Wireless and
  Microwave Technology Conference (WAMICON)}, 2016, pp. 1--5.

\bibitem{3GPP_38_901}
3rd Generation Partnership Project~(3GPP), ``{Technical Specification Group
  Radio Access Network; Study on channel model for frequencies from 0.5 to 100
  GHz (Rel-16)},'' Technical Report 38.901, ver 16.1.0, Dec. 2019.

\bibitem{chen2009multi}
M.~Chen, X.~Liang, V.~Leung, and I.~Balasingham, ``Multi-hop mesh cooperative
  structure based data dissemination for wireless sensor networks,'' in
  \emph{11th International Conference on Advanced Communication Technology},
  vol.~1.\hskip 1em plus 0.5em minus 0.4em\relax IEEE, 2009, pp. 102--106.

\end{thebibliography}

\end{document}